\documentclass[a4paper,11pt]{article}
\usepackage{jheppub} 
\usepackage{lineno}

\usepackage{graphicx}
\usepackage{dcolumn}
\usepackage{bm}

\usepackage{tikz}
\usepackage{tikz-feynman}
\usetikzlibrary{decorations.markings,decorations.pathmorphing,shapes}
\usetikzlibrary{angles,quotes} 
\usetikzlibrary{arrows.meta} 
\tikzstyle{photon}=[line width=0.4,decorate, decoration={snake,amplitude=2,segment length=4,post length=0.1}]
\usepackage[italicdiff]{physics}
\usepackage[normalem]{ulem}

\begin{document}

\title{Influence of a perfectly conducting plate on the Uehling potential of QED }

\author[a,b]{T. Azevedo,}
\author[c]{F.A. Barone,}
\author[a]{C. Farina,}
\author[d]{R. de Melo e Souza}
\author[a]{and G. Zarpelon}
\affiliation[a]{Instituto de F\'{i}sica, Universidade
Federal do Rio de Janeiro, \\
Caixa Postal 68528, RJ 21941-972, Brazil}       
\affiliation[b]{Institute of Physics of the Czech Academy of Sciences \& CEICO,\\
  Na Slovance 2, 182 21, Prague -- Czechia}
\affiliation[c]{IFQ - Universidade Federal de Itajubá, Av. BPS 1303, Pinheirinho, Caixa Postal 50, 37500-903, Itajubá, MG -- Brazil} 
\affiliation[d]{
 Instituto de F\'isica, Universidade Federal Fluminense, Avenida Litor\^anea s/n, CP 24210-346 Niter\'oi,
Rio de Janeiro -- Brazil
} 

\emailAdd{thales@.if.ufrj.br}
\emailAdd{fbarone@unifei.edu.br}
\emailAdd{farina@if.ufrj.br}
\emailAdd{reinaldos@id.uff.br}
\emailAdd{gzarpelon0@gmail.com}

\abstract{In this work, we investigate the influence of a perfectly conducting plate on the Uehling potential of Quantum Electrodynamics (QED), corresponding to the first loop correction to the classical Coulomb potential in that situation. We use the method of images adapted to the photon propagator, extending the method beyond the standard (classical) tree level calculation. We show that the effect of the plate on the quantum correction is much stronger than the expectation from a naive application of the method of images.}


\maketitle

\flushbottom


\section{\label{sec:level1}Introduction and Overview}

Classical electromagnetism is a very solid and robust theory that took many centuries to be conceived and put into solid grounds. Its construction was a big enterprise made not only by a few scientists, but also by a lot of brilliant minds who contributed since the ancient Greece with their experimental observations and theoretical developments. We can say that the creation of classical electromagnetism culminated with Maxwell’s equations, established in the 19th century, which synthesized the dynamical laws governing the time evolution of the electromagnetic fields, at least for prescribed sources (in the general case, of course, the dynamical evolution of the sources also have  to be considered). One of the main features of Maxwell’s electromagnetic theory is its linearity. In other words, superposition principle is valid. In our day by day life we benefit from this property (the huge amount of electromagnetic waves arriving at our mobile phones, for instance, do not interact with each other). Nonetheless, one must keep in mind that propagation of electromagnetic waves in continuous media can display important nonlinear features due to the induction of charge densities and electric currents. 

    In spite of the success of classical electrodynamics, up to now, the ultimate theory that describes the radiation-matter interaction is quantum electrodynamics (QED), which is considered one of the most precise theories that have ever been created by the human intellect (if not the most), with an astonishing agreement between theoretical predictions and experimental data. Maxwell's equations remain valid, although now the fields must be taken as operators instead of classical vector fields. Quantum effects manifest themselves through nontrivial properties of quantum vacuum, specially continuous creation and annihilation of virtual particle-anti-particle pairs. This means that the QED vacuum behaves as a nonlinear material which can be polarized, leading to nonlinear phenomena which can be described by effective Lagrangians~\cite{dittrich}. As a first example, we mention Delbruck scattering~\cite{meitner1933}, which consists of a process in which a photon is elastically scattered by the Coulomb field of an atomic nucleus. This is a remarkable effect, since the photon has no electric charge, and it has already been observed experimentally~\cite{jarlskog1973,muckenheim1980,rullhusen1981}. Photon splitting is another surprising effect that can occur due to vacuum polarization  and has also been observed experimentally~\cite{di2007photon}.
    
Another surprising effect is light-by-light scattering, first investigated  by Hans Euler in his PhD thesis, following a suggestion given by Peter Debye to his advisor Heisenberg~\cite{heisenbergeuler1936}---see also Refs.~\cite{Weisskopf:1936hya,Schwinger}. This phenomenon had to wait approximately 80 years to be observed \cite{atlas1,atlas2}.  Finally, we mention the correction to the Coulomb potential created by a point charge at rest due to vacuum polarization effect, usually referred to as the Uehling potential~\cite{uehling1935}. 
Since its original proposition, this potential has been extensively studied~\cite{wichmann1956,huang1976,Petelenz1987,frolov2012,Indelicato2014,greiner}, not only because of its intrinsic importance, but also since its knowledge is necessary in many situations, as for instance in the calculation of the Lamb shift~\cite{lamb1947,bethe1947,karshenboim1999polarization,yerokhin2019theory,Frugiuelea2022, Krachkov2023}.

The Uehling potential still remains an active focus of research. Different representations and calculation schemes have been proposed~\cite{Burgess2017,barone2018,Frolov2021,Mohr2023,Frolov2024}, and higher-order corrections have been investigated~\cite{Ivanov2024, Flynn2025b}. Generalizations of Uehling's result to modified QED theories have been analyzed~ \cite{Ajamieh2024, Ajamieh2025}, as well as extensions to the hadronic case~\cite{breidenbach2022hadronic,Dizer2023} and to quantum gravity theories~\cite{Draper2020,Cordova2022,Jimu2025}. In addition, the Uehling potential has played a central role in the analysis of precise experimental tests for QED~\cite{Sanamyan2023,Maza2025} and in scattering phenomena~\cite{Amundsen2021,Xu2022,Amundsen2024,Fontes2025}. Uehling's correction has also been applied for a deep understanding of the spectra of many-electron atoms~\cite{Kumar2022,Fairhall2023,Hasted2025}, ordinary molecules~\cite{Janke2025,Flynn2025}, and muonic and pionic molecules~\cite{Karr2013,Michel2019}. 

Thus far, there have been no studies  concerning the influence of boundaries on the Uehling correction. However, it is well known that the presence of objects can strongly affect the quantum vacuum. One of the most emblematic effects of this kind is the Casimir effect, which consists in its simplest form of a finite shift of the zero-point energy of the (quantized) electromagnetic field, giving rise to an attractive force between two parallel, perfectly conducting   plates~\cite{casimir1948, MiltonBook, BordagBook, Elizalde1991, farina2006casimir}.

In this work, we fill this gap by considering the effect of Dirichlet boundary conditions, realized by a perfectly conducting plate, on the first loop correction to Coulomb's law. We show that there are regions in space where the Uehling correction can be enhanced by orders of magnitude due to non-additivity effects. We also show that a naive application of the method of images misses the non-linear contribution and grossly underestimates the influence of the boundary.

The remainder of the paper is organized as follows. In Section~\ref{section:standard_uehling}, we review the standard derivation of the Uehling potential. Then, in Section~\ref{section:plate},  we use the method of images to study the first loop correction to Coulomb's law and obtain the analogue of the Uehling potential when a perfectly conducting plate is present. Some numerical results, along with a discussion on their interpretation, are presented in section~\ref{section:results}.
Finally, we conclude with a few comments in section~\ref{secion:conclusion}. We adopt the mostly negative metric signature, $\eta_{\alpha\beta} = \text{diag}(1,-1,-1,-1)$, with indices running from $0$ to $3$ (where $x^3$ corresponds to the $Z$ axis component) and express three-vectors in bold.

\section{The Standard Uehling Potential \label{section:standard_uehling}}

\subsection{Preliminaries}

The standard calculation of the Uehling Potential in QED can be found in multiple different sources \cite{peskin, greiner, schwartz2013}, but we outline the procedure to obtain such a potential  in order to use many of the results later. 

The basic Lagrangian for QED, $\mathcal{L}_{\text{QED}}$, is given by
\begin{eqnarray}
    \mathcal{L}_{\text{QED}} = -\frac{1}{4}F_{\mu\nu}F^{\mu\nu} - \overline{\Psi}(i\gamma^\mu D_\mu - m)\Psi \,.\label{QED_Lagrangian}
\end{eqnarray}
Here $m$ is the fermion mass, $F_{\mu\nu} = \partial_\mu A_\nu - \partial_\nu A_\mu$, $A_\mu$ is the photon field, $\Psi$ corresponds to the fermionic field, with $\overline{\Psi} = \gamma^0 \Psi$ and $D_\mu:= \partial_\mu - ieA_\mu$ corresponds to the covariant derivative. By use of the functional generator of correlation functions, one can construct the usual Feynman diagrammatic perturbation series in terms of the (bare) charge $e$. 

Of especial importance to us are the photon and fermion propagators, $\Delta_{\mu\nu}$ and $G_F$. In order to keep track of the loop order, we will add the $(0)$ superscript to denote the tree level (free) propagator, and $(1)$ to denote the 1 loop corrected propagator. Since we are  concerned with photon exchange processes, we will not consider loop corrections to the fermion propagators $G_F$. Keeping these conventions in mind, the explicit forms of the propagators in momentum space are simply 
\begin{eqnarray}
    \Delta^{(0)}_{\mu\nu}(p) =&& \frac{-i}{p^2 + i\epsilon}\left[\eta_{\mu\nu} - \left( 1 - \xi\right)\frac{p_\mu p_\nu}{p^2}\right],\label{photon1}\\
    G_F(p) = && \frac{i(\gamma^\mu p_\mu + m)}{p^2 - m^2 + i\epsilon}\,,\label{fermion1}
\end{eqnarray}
where $\xi$ is the gauge parameter and $\epsilon$ is the standard parameter associated with the Feynman prescription for the propagators.

The one-loop corrected photon propagator $\Delta^{(1)}_{\mu\nu}(p)$ can be represented by the leading term in $\alpha$, the fine structure constant. Diagrammatically, one has the following representation:
	\begin{align}
		 \Delta^{(1)}_{\mu\nu}(p) \ =&\  \vcenter{\hbox{\begin{tikzpicture}[scale=1]
	 	\begin{feynman}
		      \vertex (a) at (-1,0){$\mu$};
		      \vertex (b) at (1,0){$\nu$};
		      \diagram* {(a) -- [photon,edge label=$p$] (b),};
		\end{feynman}
		\end{tikzpicture}}}\ +\  \vcenter{\hbox{\begin{tikzpicture}
	 	\begin{feynman}
		      \vertex (a) at (-1,0){$\mu$};
		      \vertex (b) at (2,0){$\nu$};
		      \vertex (bob1) at (0,0);
		      \vertex (bob2) at (1,0);
		      \diagram* {(a) -- [photon,edge label=$p$] (bob1) -- [fermion, half left, looseness=1.5, edge label=\(k\)](bob2),
	       			 (bob2) -- [fermion, half left, looseness=1.5, edge label=\(k-p\)](bob1),
	       			 (bob2) -- [photon, edge label = $p$] (b),};
		\end{feynman}
		\end{tikzpicture}}}=\, \Delta^{(0)}_{\mu\nu} + \Delta^{(0)}_{\mu\rho} i\Pi^{\rho\sigma}  \Delta^{(0)}_{\sigma\nu}\,.
		\label{Propagator1}
	\end{align}

\noindent The $\Pi^{\rho\sigma}(p)$ is known as the \textit{polarization tensor} and encodes the first loop correction to the propagator.

\subsection{Fermionic Loop \label{section:fermion_loop}}

As discussed above, the fermion loop is the first loop correction to the free-photon propagator. The evaluation of such a loop can be found in multiple different sources---see, for example, Refs.~\cite{greiner,peskin}. By using dimensional regularization with $d = 4 - \epsilon$, one can isolate the divergence of the loop and write the polarization tensor as
\begin{eqnarray}
\Pi^{(1)\mu\nu}(p) =-\frac{e^2}{2\pi^2} \left(\eta^{\mu\nu} p^2 -  p^\mu p^\nu\right)    \phantom{\int_0^1 d{x}\, \left(\frac{2}{\epsilon} -\gamma_{E}-\frac{1}{2} + \ln\left[\frac{4\pi\mu^2}{m^2 + p^2 x(x-1)}\right] \right)}\nonumber\\
\times \int_0^1 d{x}\, x(1-x)\left(\frac{2}{\epsilon} -\gamma_{E}-\frac{1}{2} + \ln\left[\frac{4\pi\mu^2}{m^2 + p^2 x(x-1)}\right] \right),
\label{Renorm:Pi4dfinal}
\end{eqnarray}
where $\gamma_{E}$ is the Euler-Mascheroni constant and $\mu$ is a mass scale introduced in dimensional regularization to ensure that the coupling constant remains dimensionless in $d$-dimensional spacetime.

Following the on-shell renormalization scheme, the divergent contribution in Eq.~(\ref{Renorm:Pi4dfinal}) is canceled by a wavefunction renormalization counterterm for the gauge field, along with a redefinition of the electric charge, $e \mapsto e_{R}$ \cite{WeinbergeQFT}. Accordingly, the one-loop renormalized vacuum polarization tensor $\Pi^{(1)\mu\nu}_R$ is introduced as
\begin{eqnarray}
    \Pi^{(1)\mu\nu}_R(p):= \frac{e_R^2}{2\pi^2}(\eta^{\mu\nu}p^2 - p^\mu p^\nu)\int_0^1 \!d{x}\, x(1-x)\ln\left[1 + \frac{p^2}{m^2}x(x-1)\right]\!.\label{Renormalized_Pi}
\end{eqnarray}
The renormalized polarization tensor given by Eq.~(\ref{Renormalized_Pi}) appears as a correction to the renormalized free propagator as can be seen by the diagrammatic equation below. 
\begin{figure}[h]
	\begin{align}
		 \Delta^{(1)}_{\mu\nu}(p) \ =&\  \vcenter{\hbox{\begin{tikzpicture}[scale=1]
	 	\begin{feynman}
		      \vertex (a) at (-1,0){$\mu$};
		      \vertex (b) at (1,0){$\nu$};
		      \diagram* {(a) -- [photon] (b),};
		\end{feynman}
		\end{tikzpicture}}}\ +\ \vcenter{\hbox{\feynmandiagram[small,horizontal=a to b]{a[particle=$\mu$] -- [photon] c [crossed dot] -- [photon]b[particle=$\nu$]} ;}} \nonumber\\
            & + \ \vcenter{\hbox{\begin{tikzpicture}
	 	\begin{feynman}
		      \vertex (a) at (-1,0){$\mu$};
		      \vertex (b) at (2,0){$\nu$};
		      \vertex (bob1) at (0,0);
		      \vertex (bob2) at (1,0);
		      \diagram* {(a) -- [photon] (bob1) -- [fermion, half left, looseness=1.5](bob2),
	       			 (bob2) -- [fermion, half left, looseness=1.5](bob1),
	       			 (bob2) -- [photon] (b),};
		\end{feynman}
		\end{tikzpicture}}} \nonumber\\
		=&\, \Delta^{(0)}_{\mu\nu} +  \Delta^{(0)}_{\mu\rho} i\Pi_R^{\rho\sigma}  \Delta^{(0)}_{\sigma\nu}\,.
		\label{Propagator2}
	\end{align}
\end{figure}

In the literature, it is common to factorize the tensorial nature of the polarization tensor from the scalar part by writing $\Pi^{(1)\mu\nu}_R(p) = (\eta^{\mu\nu}p^2 -p^\mu p^\nu) \Pi_R(p^2)$. This is useful because it allows one to factor the loop-corrected propagator as 
\begin{eqnarray}
    \Delta^{(1)}_{\mu\nu}(p) = \frac{-i\eta_{\mu\nu}}{p^2 + i\epsilon}\left[1 + \Pi_R(p^2)\right] + (p_\mu p_\nu\text{ terms})\,.
\end{eqnarray}

The $p_\mu p_\nu$ terms above vanish when one couples the propagator to a static charge, but are known to cause gauge ambiguities for moving sources \cite{Hida_Okamura10.1143/PTP.47.1743}. Since we are interested in coupling to static sources, it is therefore sufficient to disregard these pieces, which is equivalent to choosing Feynman gauge ($\xi =1$) and focusing on the terms proportional to $\eta_{\mu\nu}$, in the propagator.

\subsection{Extraction of the potential \label{section:extraction_potential}}

To find the effect of this correction on the potential created by a given charge distribution, one can use the corrected propagator as a Green's function to Maxwell's inhomogeneous equation, that is\footnote{Note that the factor of $i$ on the left-hand side of Eq.~(\ref{potential1}) is needed for consistency, since the photon propagator also contains a factor of $i$. Otherwise the resulting potential would be purely imaginary.}
\begin{eqnarray}
    iA^{(1)}_\mu(x) = \int\dd[4]{x'} \Delta^{(1)}_{\mu\nu}(x - x') j^\nu(x')\,.\label{potential1}
\end{eqnarray}
In Feynman gauge, the loop corrected photon propagator in position space is the (inverse) Fourier transform of its momentum representation, that is
\begin{eqnarray}
    \Delta^{(1)}_{\mu\nu}(x-x') = \int\frac{\dd[4]{p}}{(2\pi)^4}\frac{-i \eta_{\mu\nu}}{p^2}\left[1 + \Pi_R(p^2)\right] e^{-i p\cdot(x - x')}\,.
    \label{1loopPropagator}
\end{eqnarray}
For the case of a static point-like charge, the current density assumes the form $j^\nu(x') = q \delta^3(\vb{x'} - \vb{x}_s) \delta^\nu_{\ 0}$, where $\vb{x}_s$ corresponds to the position of the charge (i.e. the source). Insertion of Eq.~(\ref{1loopPropagator}) into Eq.~(\ref{potential1}) with this static point current $j^\nu$ yields $A_\mu^{(1)}=\eta_{\mu0}\phi^{(1)}$ with
\begin{eqnarray}
    \phi^{(1)}(\vb{x}) = \phi^{(0)}(\vb{x}) + q \int\frac{\dd[3]{\vb{p}}}{(2\pi)^3} \frac{\Pi_R(-\vb{p}^2)}{\vb{p}^2}  e^{i\vb{p}\cdot(\vb{x}-\vb{x}_s)}\,,
    \label{Eq10}
\end{eqnarray}
where $\phi^{(0)}(\vb{x})$ corresponds to the standard Coulomb potential, i.e. $\phi^{(0)}(\vb{x}) := q/4\pi|\vb{x}-\vb{x}_s|$. The second term corresponds to the loop contribution and can be analytically computed. For this, it is common to rewrite the expression for the polarization scalar $\Pi_R(p^2)$. Following \cite{beretetskii,greiner}, one integrates by parts in order to remove the logarithm and, upon a change of variables of the form $v = 2x -1$, writes 
\begin{eqnarray}
    \Pi_R(-\vb{p}^2) = \frac{\alpha}{\pi} \frac{\vb{p}^2}{4m^2}\int_{0}^1 \dd{v} \frac{v^2\left(1 - \frac{1}{3}v^2\right)}{1 + (\vb{p}/2m)^2(1-v^2)}\,,\label{param_Pi2}
\end{eqnarray} 
where $\alpha = e_R^2/4\pi$ is the fine structure constant. Inserting (\ref{param_Pi2}) into Eq.~(\ref{Eq10}) and exchanging the order of integration, one has
\begin{eqnarray}
    \phi^{(1)}(r) = \phi^{(0)}(r) +  \frac{q}{4\pi r} \frac{\alpha}{\pi} \int_0^1\dd{v} \frac{v^2(1 - \frac{1}{3}v^2)}{1 - v^2} e^{-\frac{2m r}{\sqrt{1-v^2}}}\,,\nonumber
\end{eqnarray}
where we have labeled $r \equiv |\vb{x} - \vb{x}_s|$ to simplify the notation. After a change of variables $u = (1-v^2)^{-1/2}$, the above equation can be cast into the standard form found in the literature, i.e.
\begin{eqnarray}
    \phi^{(1)}(r) = \frac{q}{4\pi r}\left[1 +\frac{2\alpha}{3\pi} \int_1^\infty\!\!\dd{u}\left(1 + \frac{1}{2u^2}\right)\frac{\sqrt{u^2 -1}}{u^2} e^{-2mur}\right]\!.
    \label{UehlingPotential1}
\end{eqnarray}
Eq.~(\ref{UehlingPotential1}) is known as the \emph{Uehling potential} of QED and, as anticipated, gives the correction of order $\alpha$ to Coulomb's law. The above can also be expressed in terms of Meijer G-functions \cite{barone2018}.

From Eq.~(\ref{UehlingPotential1}), one can also analyze the limiting behaviors for $r\gg1/m$ and $r\ll1/m$. For $r\gg1/m$, the potential can be approximated as \cite{greiner}
\begin{eqnarray}
    \phi^{(1)}(r) \approx \frac{q}{4\pi r}\left[1 + \frac{\alpha}{4\sqrt{\pi}}\frac{e^{-2mr}}{(mr)^{3/2}} \right]\,.
\end{eqnarray}
For $r\ll1/m$, one finds a logarithmic divergence which yields
\begin{eqnarray}
   \phi^{(1)}(r)\approx \frac{q}{4\pi r}\left[1 + \frac{2\alpha}{3\pi}\left(\ln\frac{1}{mr} - \frac{5}{6} - \gamma_E\right)\right]\,,
\end{eqnarray}
 In both limits, the interpretation of the deviations lead to the notion of a \textit{running coupling constant} $e^{(1)}(r):=4\pi r\phi^{(1)}(r)$ which effectively alters the charge ``felt" by a probe near or far from the source.

\section{Influence of a Perfectly Conducting Plate}\label{section:plate}

 Let us now consider what happens to the Uehling potential once a perfectly conducting plate is placed at the origin $z=x^3 =0$. To do so, we first employ the image method to obtain the propagator.

\subsection{The method of images for propagators \label{method_of_images}}

The method of images is a technique for calculating the electrostatic potential of a given charge distribution in the presence of Dirichlet boundary conditions. 
%
%
%
Usually, the method is introduced to deal with one of the simplest kinds of boundaries, namely a perfectly conducting infinite plate. In that case, the (classical) solution is directly obtained as a sum of two contributions, one consisting of the potential generated by the real charge and another generated by an image with the opposite charge placed at the reflected coordinates with respect to the plane of plate. 

Interestingly, the method of images can also be formulated in terms of  propagators~\cite{Brown1969,de1997image}---see also Ref.~\cite{Polchinski1} for a similar discussion in the context of open strings. 
To illustrate this, let us consider how to re-obtain the standard result for the potential generated by a charge $e$ next to a perfectly conducting plate placed at $z = x^3 = 0$. In this case, the electromagnetic potential must satisfy the Dirichlet boundary conditions at $x^3 = 0$. At the propagator level, this implies that the \emph{photon} propagator must be modified as
\begin{eqnarray}
    {\overline{\Delta}^{(0)}_{\mu\nu}(x,y)}_{\tilde{y}} = \Delta^{(0)}_{\mu\nu}(x - y) - \Delta^{(0)}_{\mu\nu}(x - \widetilde{y})\,,\label{modified_propagator_prescription1}
\end{eqnarray}
where $\widetilde{y}^\mu = (y^0, y^1, y^2, -y^3)$. It is readily verified that the modified propagator satisfies the boundary conditions at $z = 0$, that is ${\overline{\Delta}^{(0)}_{\mu\nu}(x^3=0)}_{\tilde{y}} = 0 = {\overline{\Delta}^{(0)}_{\mu\nu}(y^3 = 0)}_{\tilde{y}}$. Note that one could equally well reflect the other coordinate of the propagator and write
\begin{eqnarray}
    {\overline{\Delta}^{(0)}_{\mu\nu}(x,y)}_{\tilde{x}} = \Delta^{(0)}_{\mu\nu}(x - y) - \Delta^{(0)}_{\mu\nu}(\widetilde{x} - y)\,,\label{modified_propagator_prescription2}
\end{eqnarray}
where, again, the tilde represents reflection of the fourth  variable ($z$) with respect to the origin. Both Eq.~(\ref{modified_propagator_prescription1}) and Eq.~(\ref{modified_propagator_prescription2}) satisfy the boundary conditions. Often, the momentum space representations of the above are needed. For the first prescription, one has
\begin{eqnarray}
    {\overline{\Delta}^{(0)}_{\mu\nu}(x,y)}_{\tilde{y}} = \int\frac{\dd[4]{p}}{(2\pi)^4} \Delta^{(0)}_{\mu\nu}(p) e^{-ip\cdot(x-y)} \chi_p(y)\,,
\end{eqnarray}
where we defined
\begin{eqnarray}
    \chi_p(y) := 1 - e^{2ip^3 y^3}\,.
\end{eqnarray}
For the second prescription, one has
\begin{eqnarray}
    {\overline{\Delta}^{(0)}_{\mu\nu}(x,y)}_{\tilde{x}} = \int\frac{\dd[4]{p}}{(2\pi)^4} \Delta^{(0)}_{\mu\nu}(p) e^{-ip\cdot(x-y)} \chi^*_p(x)\,,
\end{eqnarray}
where now $\chi^*_p(x) = \chi_p(-x) = 1 -e^{-2ip^3 x^3}$.

Here it is useful to emphasize the following property which relates the two prescriptions. If we note that $p\cdot\widetilde{x} = \widetilde{p} \cdot x$, it is clear that 
\begin{eqnarray}
    \int\dd[4]{p} \Delta^{(0)}_{\mu\nu}(p) e^{-ip\cdot(x - \widetilde{x}')} = \int\dd[4]p \Delta^{(0)}_{\mu\nu}(\widetilde{p}) e^{-ip\cdot(\widetilde{x} - x')}\,.~~~~~~~ 
\end{eqnarray}
Thus, if $\Delta^{(0)}_{\mu\nu}(\widetilde{p}) = \Delta^{(0)}_{\mu\nu}(p)$, \textit{both prescriptions are the same}. Since this is true in Feynman gauge and we will be working with this choice, we will choose the prescription that simplifies calculations.

As a proof of concept, let us consider the standard Coulomb potential generated by the static charge $q$ placed at $\vb{x}_s = (x_s^1,x_s^2,L)$. To obtain the potential, one can follow the procedure outlined in  section \ref{section:extraction_potential}. Namely one writes the modified potential as
\begin{eqnarray}
    i\overline{A}^{(0)}_\mu(x)  = \int_{\mathbb{R}^{2,1}\times\mathbb{R}_+}\dd[4]{y} \overline{\Delta}^{(0)}_{\mu\nu}(x, y) j^\nu(y)\,,\label{modified_standardpotential}
\end{eqnarray}
for $x^3\geq 0$, while $\overline{A}^{(0)}_\mu(x) = 0$, for $x^3<0$. As before, the static charge is modeled by $j^\nu(y) = q \delta^{\nu}_{\ 0} \delta^{3}(\vb{y} - \vb{x}_s)$ so that we have again $\overline{A}_\mu^{(0)}(x) =\eta_{\mu0}\overline{\phi}^{(0)}(x)$. By using Feynman gauge 
and inserting this current density into Eq.~(\ref{modified_standardpotential}), with either prescription (\ref{modified_propagator_prescription1}) or (\ref{modified_propagator_prescription2}), one finds the usual result from electrostatics, i.e.
\begin{eqnarray}
    \overline{\phi}^{(0)}(x) = \frac{1}{4\pi}\left(\frac{q}{|\vb{x} - \vb{x}_s|} - \frac{q}{|\vb{x} - \widetilde{\vb{x}}_s|}\right)\,,\label{tree_potential}
\end{eqnarray}
where $\widetilde{\vb{x}}_s := (x_s, y_s, -L)$. This allows for the interpretation that the potential can be modeled by the standard single Coulomb potential with an added contribution coming from a fictitious image with opposite charge and placed at the reflected position of the real charge. Now we are ready to move on to the first loop level contribution.

\subsection{Loop corrected propagator}

Given the photon propagator modified by the presence of the plate, one can ask what happens with the Uehling potential. In particular, can it still be interpreted as the potential without the plate plus a contribution from an image charge? 

Following the procedure outlined above, one must first calculate the modified, 1 loop corrected photon propagator $\overline{\Delta}^{(1)}_{\mu\nu}$, which corresponds to a sum of the tree-level modified propagator $\overline{\Delta}^{(0)}_{\mu\nu}$ and the loop contribution $\delta\overline{\Delta}^{(1)}_{\mu\nu}$. Coloring the modified photon propagator lines in red, we can represent the calculation diagrammatically by (see Fig.~\ref{diagram_prescription} for a more precise representation of $\delta\overline{\Delta}^{(1)}_{\mu\nu}$ in position space)


\begin{figure}[h]
	\begin{align}
		 \overline{\Delta}^{(1)}_{\mu\nu}(p) \ =&\ \ \vcenter{\hbox{\begin{tikzpicture}[scale=1]
	 	\begin{feynman}
		      \vertex (a) at (-1,0){$\mu$};
		      \vertex (b) at (1,0){$\nu$};
		      \diagram* {(a) -- [photon, red] (b),};
		\end{feynman}
		\end{tikzpicture}}}\ +\ \ \vcenter{\hbox{\begin{tikzpicture}
	 	\begin{feynman}
		      \vertex (a) at (-1,0){$\mu$};
		      \vertex (b) at (2,0){$\nu$};
		      \vertex (bob1) at (0,0);
		      \vertex (bob2) at (1,0);
		      \diagram* {(a) -- [photon, red] (bob1) -- [fermion, half left, looseness=1.5, edge label=\(k\)](bob2),
	       			 (bob2) -- [fermion, half left, looseness=1.5, edge label=\(k-p\)](bob1),
	       			 (bob2) -- [photon, red] (b),};
		\end{feynman}
		\end{tikzpicture}}}\nonumber\\
            =&\ \overline{\Delta}^{(0)}_{\mu\nu} + \overbrace{\overline{\Delta}^{(0)}_{\mu\alpha}(p) i\Pi^{\alpha\beta}(p) \overline{\Delta}^{(0)}_{\beta\nu}(p)}^{\delta\overline{\Delta}^{(1)}_{\mu\nu}}\,.
		\label{dressedprop2}
	\end{align}
\end{figure}

This pure loop contribution $\delta\overline{\Delta}^{(1)}_{\mu\nu}$ can be obtained by the standard methods and, in position space, we have
\begin{eqnarray}
    \delta\overline{\Delta}^{(1)}_{\mu\nu}  &=& ie^2\int_{\mathcal{D}}\dd[4]{x_1} \int_{\mathcal{D}}\dd[4]{x_2} \overline{\Delta}^{(0)}_{\mu\alpha}(x,x_1) \gamma^\alpha G_F(x_1 - x_2) \gamma^\beta G_F(x_2 - x_1) \overline{\Delta}^{(0)}_{\beta\nu}(x_2 , y)\nonumber\\
    &=& \int_{\mathcal{D}} d^4x_1 \int_{\mathcal{D}} d^4x_2 \overline{\Delta}^{(0)}_{\mu\alpha}(x,x_1) i\Pi^{\alpha\beta}(x_1, x_2) \overline{\Delta}_{\beta\nu}^{(0)}(x_2,y)\,,
    \label{1plate:modified_loop_contribution1}
\end{eqnarray}
where $\Pi^{\alpha\beta}$ represents the loop part of the expression and $\mathcal{D} = \mathbb{R}^{2,1}\times\mathbb{R}_+$ is the domain of integration with only the positive semi-axis for the variable perpendicular to the plate (i.e. $x^3 = z$). It is noteworthy that, by construction, this correction satisfies the Dirichlet boundary conditions. 

Evaluation of expression (\ref{1plate:modified_loop_contribution1}) can be simplified if we take prescription (\ref{modified_propagator_prescription1}) for the second propagator and prescription (\ref{modified_propagator_prescription2}) for the first. With this choice, the momentum space representation of Eq.~(\ref{1plate:modified_loop_contribution1}) can be written as
\begin{eqnarray}
    \delta\overline{\Delta}^{(1)}_{\mu\nu}  &=&  \int\left(\prod_{i=1}^{4}\frac{\dd[4]{p_i}}{(2\pi)^4}\right)\int_{\mathcal{D}}\dd[4]{x_1}\int_{\mathcal{D}}\dd[4]{x_2}\chi_{p_1}^*(x^3)\chi_{p_4}(y^3) \nonumber\\
    &&\times \Delta^{(0)}_{\mu\alpha}(p_1)  i\Pi^{\alpha\beta}(p_2, p_3)  \Delta^{(0)}_{\beta\nu}(p_4) e^{-ip_1\cdot(x-x_1)} \nonumber\\
    &&\times e^{-ip_2\cdot(x_1 - x_2)} e^{-ip_3\cdot(x_2 - x_1)}e^{-ip_4\cdot(x_2- y)}\,.\label{1plate:deltaDelta1.1}
\end{eqnarray}
Now, one can gather the $x_1$ and $x_2$ dependent phases and integrate the above expression. However, since the domain of integration $\mathcal{D}$ does not correspond to the full $\mathbb{R}^{3,1}$ space, the integration over the half-axis must be carefully treated.

As explained in  Appendix \ref{App_A}, use of the Sokhotski-Plemelj theorem produces four terms which we will name $a_{\mu\nu}, b_{\mu\nu}, c_{\mu\nu}, d_{\mu\nu}$. 
The first, $a_{\mu\nu}$, is the simplest, since it involves the full momentum delta product $\delta^4(p_1 - p_2 + p_3)\delta^4(p_2 - p_3 - p_4)$. Therefore, after changing variables to $p_2 = k$, $p_3 - p_2 = q$, and integrating with the deltas, it reduces to 
\begin{eqnarray}
    a_{\mu\nu} &=& \frac{1}{4}\int\frac{\dd[4]{p_1}}{(2\pi)^4}\int\frac{\dd[4]{k}}{(2\pi)^4}\Delta^{(0)}_{\mu\alpha}(p_1)  i\Pi^{\alpha\beta}(k, k-p_1)  \Delta^{(0)}_{\beta\nu}(p_1)\nonumber\\
    &&\times\chi_{p_1}^*(x^3)\chi_{p_1}(y^3) e^{-ip_1\cdot (x-y)}\,,\label{1plate:Omega1}
\end{eqnarray}
which, after relabeling $p_1 \to p$ and redefining the tensor $\Pi^{\alpha\beta}$ to absorb the $k$ integral, becomes
\begin{eqnarray}
    a_{\mu\nu} &=& \frac{1}{4}\int\frac{\dd[4]{p}}{(2\pi)^4} \Delta^{(0)}_{\mu\alpha}(p)  i\Pi^{\alpha\beta}(p)  \Delta^{(0)}_{\beta\nu}(p)\chi_{p}^*(x^3)\chi_p(y^3) e^{-ip\cdot (x-y)}\,.
    \label{1plate:Omega1.2}
\end{eqnarray}

The second and third terms arising from expression (\ref{1plate:deltaDelta1.1}), i.e. $b_{\mu\nu}$ and $c_{\mu\nu}$, are similar. For $b_{\mu\nu}$, one has
\begin{eqnarray}
    b_{\mu\nu} &=& \int\frac{\dd[4]{p_1}}{(2\pi)^4} \int\frac{\dd[4]{k}}{(2\pi)^4} \Delta^{(0)}_{\mu\alpha}(p_1) \chi^*_{p_1}(x^3) i\Pi^{\alpha\beta}(k,k-p_1)  \nonumber\\
    &&\times e^{-ip_1\cdot x}\frac{i}{4\pi} \mathcal{P}\int\dd[4]{p_4} \Delta^{(0)}_{\beta\nu}(p_4) \chi_{p_4}(y^3) e^{ip_4 \cdot y}\nonumber\\
    &&\times\frac{\delta^2(\vb{p}_{1\parallel} - \vb{p}_{4\parallel}) \delta(p_1^0 - p_4^0)}{p_4^3 - p_1^3}\,,
    \label{1plate:Omega2.1}
\end{eqnarray}
where $\mathcal{P}$ refers to the principal value of the integral and we have again changed the momentum variables, with $p_2 = k$, $p_3 - p_2 = q$. The same considerations for $c_{\mu\nu}$ yield
\begin{eqnarray}
    c_{\mu\nu} &=& \int\frac{\dd[4]{p_4}}{(2\pi)^4} \int\frac{\dd[4]{k}}{(2\pi)^4} \Delta^{(0)}_{\beta\nu}(p_4) \chi_{p_4}(y^3) i\Pi^{\alpha\beta}(k,k-p_4) \nonumber\\
    &&\times  e^{ip_4\cdot y}\frac{i}{4\pi} \mathcal{P}\int\dd[4]{p_1} \Delta^{(0)}_{\mu\alpha}(p_1) \chi^*_{p_1}(x^3) e^{-ip_1 \cdot x}\nonumber\\
    &&\times\frac{\delta^2(\vb{p}_{4\parallel} - \vb{p}_{1\parallel}) \delta(p_4^0 - p_1^0)}{p_4^3 - p_1^3}\,.
    \label{1plate:Omega3.1}
\end{eqnarray}

Let us consider expression (\ref{1plate:Omega2.1}) in more detail. Using Feynman gauge and defining $\omega^2_{1\parallel} := (p_1^0)^2 - \vb{p}_{1\parallel}^2$, we can write $b_{\mu\nu}$ as
\begin{eqnarray}
     b_{\mu\nu} &=& \int\frac{\dd[4]{p_1}}{(2\pi)^4} \int\frac{\dd[4]{k}}{(2\pi)^4} \Delta^{(0)}_{\mu\alpha}(p_1) \chi^*_{p_1}(x^3) i\Pi^{\alpha\beta}(k,k-p_1)\nonumber\\
    &&\times\frac{i}{2\pi}\eta_{\beta\nu} e^{i(p^0_1 y^0 - \vb{p}_{1\parallel}\cdot \vb{y}_{\parallel}) -ip_1\cdot x}\nonumber\\
    &&\times\mathcal{P}\int_{-\infty}^{\infty}\dd{\xi} \frac{\sin(\xi y^3)}{(\xi^2 - \omega_{1\parallel}^2 - i\epsilon)\left(\xi - p_1^3\right)}\,,
    \label{eq4.40}
\end{eqnarray}
where we have relabeled $\xi = p^3_4$. The $\xi$ integral appearing above will also appear for $c_{\mu\nu}$ and, as we will see later, for $d_{\mu\nu}$. It can be explicitly evaluated by separating the sine into complex exponentials and using contour integration on each piece. Using the resulting expression, we write
\begin{eqnarray}
    b_{\mu\nu} 
    &=& \int\frac{\dd[4]{p}}{(2\pi)^4} \chi^*_{p}(x^3)  \Delta^{(0)}_{\mu\alpha}(p)  i\Pi^{\alpha\beta}(p) B_{\beta\nu}(p;y^\perp) e^{-ip\cdot(x-y)}\,,\label{1plate:Omega_2final}
\end{eqnarray}
where we have relabeled $p_1 \to p$, absorbed the resulting $k$ integral into the polarization tensor and defined
\begin{eqnarray}
    B_{\beta\nu}(p;y^\perp) := \frac{i\eta_{\beta\nu}}{2}\frac{e^{ip^\perp y^\perp}}{\omega_\parallel^2 - (p^\perp)^2} \left(e^{i\omega_{\parallel} y^\perp} - \cos(p^\perp y^\perp)\right) \,.\label{1plate:B_term}
\end{eqnarray}
Retracing the same steps for $ c_{\mu\nu}$, one can write 
\begin{eqnarray}
    c_{\mu\nu} &=& \int\frac{\dd[4]{p}}{(2\pi)^4} C_{\mu\alpha}(p;x^3) i\Pi^{\alpha\beta}(p) \Delta^{(0)}_{\beta\nu}(p) \chi_p(y) e^{-ip\cdot(x-y)}\,,\label{1plate:Omega_3final}
\end{eqnarray}
with
\begin{eqnarray}
    C_{\mu\alpha}(p;x^\perp) :=\frac{i\eta_{\mu\alpha}}{2}\frac{e^{-ix^\perp p^\perp}}{\omega_{\parallel}^2 - (p^\perp)^2} \left(e^{i\omega_{\parallel} x^\perp} - \cos(p^\perp x^\perp)\right)\,.
    \label{1plate:C_term}
\end{eqnarray}

The fourth contribution, $d_{\mu\nu}$, can be written in a similar way, but with the added complication that the polarization tensor $\Pi^{\alpha\beta}$ will appear nested inside one of the momentum integrals. However, because of the way in which $\Pi^{\alpha\beta}$ depends on the momentum, one can interchange the order of integrations to write
\begin{eqnarray}
    d_{\mu\nu} &=&\int_{\mathbb{R}^{2,1}}\frac{\dd{p_1^0}\dd[2]{p^{\parallel}_1}}{(2\pi)^6}  e^{-i\overline{p}_1\cdot(\overline{x} - \overline{y})} \int_{-\infty}^\infty \dd{p^3} i\Pi^{\alpha\beta}(\overline{p}_1, p^3)\nonumber\\
    &&\times  \mathcal{P}\int_{-\infty}^{\infty} \dd{p_1^3} \frac{\Delta^{(0)}_{\mu\alpha}(\overline{p}_1,p^3_1)}{p^3 - p_1^3} \chi^*_{p_1}(x^3) e^{ip_1^3 x^3} \nonumber\\
    &&\times \mathcal{P}\int_{-\infty}^{\infty} \dd{p^3_4} \frac{\Delta^{(0)}_{\beta\nu}(\overline{p}_1, p_4^3)}{p^3 - p_4^3} \chi_{p^3_4}(y^3) e^{-ip_4^3 y^3}\,.
    \label{1plate:Omega_4.1}
\end{eqnarray}
Here $\overline{p} := (p^0, \vb{p}^\parallel,0)$ is the vector with all of the components of the four vector $p^\mu$ except $p^3$, such that $\overline{p}^2 := (p^0)^2 - (\vb{p}^\parallel)^2$. Thus, $|\overline{p}| = \sqrt{(p^0)^2 - (\vb{p}^\parallel)^2} = \omega_{p\parallel}$. 

The two principal value integrals appearing in (\ref{1plate:Omega_4.1}) can be written as the same integral which appeared before for $b_{\mu\nu}$ and $c_{\mu\nu}$. Evaluating them in the same manner yields 
\begin{eqnarray}
    d_{\mu\nu} = \int\frac{\dd[4]{p}}{(2\pi)^4} D_{\mu\alpha\beta\nu}(p;x^3,y^3) i\Pi^{\alpha\beta}(p) e^{-ip\cdot(x-y)}\,,
    \label{1plate:Omega_4final}
\end{eqnarray}
where
\begin{align}
    D_{\mu\alpha\beta\nu}(p;x^\perp,y^\perp) :=-\frac{\eta_{\mu\alpha}\eta_{\beta\nu}}{(\omega_{\parallel}^2 - (p^\perp)^2)^2}\Big[&e^{i\omega_{\parallel}(x^\perp + y^\perp\!)}- e^{i\omega_{\parallel} y^\perp}\!\!\cos(p^\perp x^\perp\!) - e^{i\omega_{\parallel} x^\perp}\!\!\cos(p^\perp y^\perp\!)\nonumber\\
     &  + \cos(p^\perp x^\perp\!)\cos(p^\perp y^\perp\!)\!\Big] e^{-ip^\perp(x^\perp - y^\perp\!)}.
     \label{1plate:D_term}
\end{align}

Putting together all contributions from Eqs. (\ref{1plate:Omega1.2}, \ref{1plate:Omega_2final}, \ref{1plate:Omega_3final}, \ref{1plate:Omega_4final}), we write the 1~loop modified propagator in momentum space as
\begin{eqnarray}
    \overline{\Delta}^{(1)}_{\mu\nu}(p) &=& \overline{\Delta}^{(0)}_{\mu\nu}(p) + i\Pi^{\alpha\beta}(p)\bigg[\frac{1}{4}\chi_{p}^*(x^3) \Delta^{(0)}_{\mu\alpha}(p) \Delta^{(0)}_{\beta\nu}(p) \chi_{p}(y^3) + D_{\mu\alpha\beta\nu}(p;x^3,y^3)  \nonumber\\
    && \,+\; \chi^*_p(x^3) \Delta^{(0)}_{\mu\alpha}(p) B_{\beta\nu}(p;y) +  C_{\mu\alpha}(p;x) \Delta^{(0)}_{\beta\nu}(p) \chi_p(y^3) \,\bigg]\,,
    \label{1plate:1loop_modifiedpropagator}
\end{eqnarray}
where each of the $B_{\mu\nu}, C_{\mu\nu}, D_{\mu\nu}$ terms are the ones in Eqs. (\ref{1plate:B_term}, \ref{1plate:C_term}, \ref{1plate:D_term}), respectively.

Since we have been able to maintain the structure of the polarization tensor $\Pi^{\alpha\beta}$, we assume that proper renormalization can still be carried out in the same way as in the case without the plate. Pragmatically speaking, we substitute $e \mapsto e_R$ and $\Pi^{\mu\nu} \mapsto \Pi^{\mu\nu}_R(p) = (\eta^{\mu\nu}p^2 - p^\mu p^\nu) \Pi_R(p^2)$ in Eq.~(\ref{1plate:1loop_modifiedpropagator}), where the renormalized scalar $\Pi_R(p^2)$ is still given by (\ref{Renormalized_Pi}).

\subsection{Diagrammatic interpretation}

Before moving on to the extraction of the potential, let us briefly give a physical interpretation behind the prescription we used and the overall final form of the propagator. 
The plate effect can be more easily understood if we go back to Eq.~(\ref{1plate:modified_loop_contribution1}) and explicitly replace each modified tree level propagator by their definitions in Eq.~(\ref{modified_propagator_prescription1}) and Eq.~(\ref{modified_propagator_prescription2}).

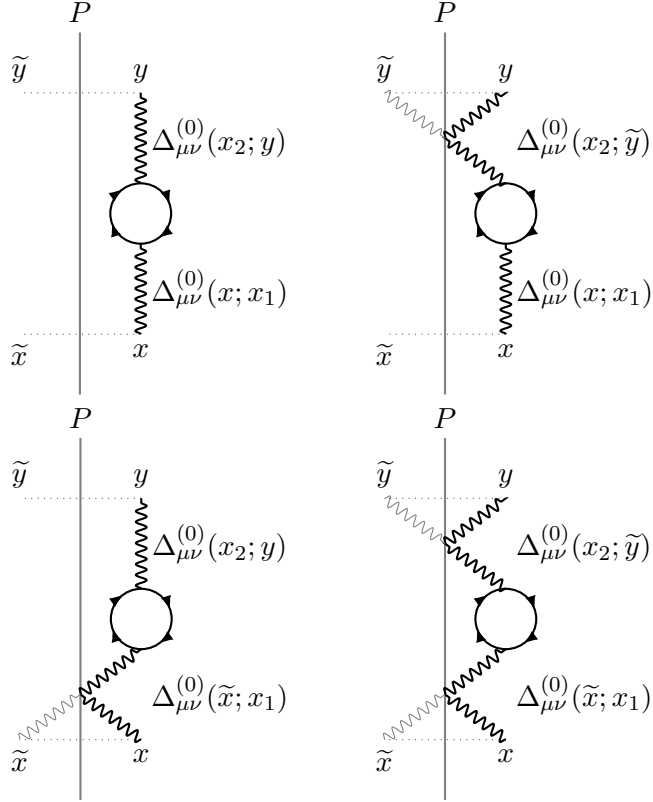
\begin{figure}[!h]
    \centering
        \begin{tikzpicture}[scale=0.8]
            \message{Attempt}
        \coordinate (plate_base) at (0,-3);
        \coordinate (plate_top) at (0,3);
        \coordinate (x) at (1,-2);
        \coordinate (y) at (1,2);
        \coordinate (x1) at (1,-0.5);
        \coordinate (x2) at (1,0.5);
        \coordinate (Rinner) at (0.5,0);
        \coordinate (Router) at (1.5,0);
        \coordinate (xi) at (-1,-2);
        \coordinate (yi) at (-1,2);
        \coordinate (r1) at (0,-1.25);
        \coordinate (r2) at (0,1.25);
        \node[above] at (plate_top) {$P$};
        \node[below] at (x) {$x$};
        \node[above] at (y) {$y$};
        \node[below] at (xi) {$\widetilde{x}$};
        \node[above] at (yi) {$\widetilde{y}$};
        \draw[-,gray,thick] (plate_base) -- (plate_top);
        \draw[photon,thick] (x) -- (x1) node[midway, right]{$\Delta^{(0)}_{\mu\nu}(x;x_1)$};
        \draw[-,black,thick,decoration={markings,mark=at position 0.2 with {\arrow{latex}}, mark=at position 0.4 with {\arrow{latex}}, mark=at position 0.7 with {\arrow{latex}},mark=at position 0.90 with {\arrow{latex}}},postaction={decorate}]
    (x1) to[out=180,in=-90] (Rinner) to[out=90,in=180] (x2) to[out=0,in=90] (Router) to[out=-90,in=0] (x1);
        \draw[photon,thick] (x2) -- (y) node[midway, right]{$\Delta^{(0)}_{\mu\nu}(x_2;y)$};
        \draw[-,dotted,opacity=0.7] (x) -- (xi);
        \draw[-,dotted,opacity=0.7] (y) -- (yi);
    \end{tikzpicture}
    \qquad 
    \begin{tikzpicture}[scale=0.8]
            \message{Attempt}
        \coordinate (plate_base) at (0,-3);
        \coordinate (plate_top) at (0,3);
        \coordinate (x) at (1,-2);
        \coordinate (y) at (1,2);
        \coordinate (x1) at (1,-0.5);
        \coordinate (x2) at (1,0.5);
        \coordinate (Rinner) at (0.5,0);
        \coordinate (Router) at (1.5,0);
        \coordinate (xi) at (-1,-2);
        \coordinate (yi) at (-1,2);
        \coordinate (r1) at (0,-1.25);
        \coordinate (r2) at (0,1.25);
        \node[above] at (plate_top) {$P$};
        \node[below] at (x) {$x$};
        \node[above] at (y) {$y$};
        \node[below] at (xi) {$\widetilde{x}$};
        \node[above] at (yi) {$\widetilde{y}$};
        \draw[-,gray,thick] (plate_base) -- (plate_top);
        \draw[photon,thick] (x) -- (x1) node[midway, right]{$\Delta^{(0)}_{\mu\nu}(x;x_1)$};
        \draw[-,black,thick,decoration={markings,mark=at position 0.2 with {\arrow{latex}}, mark=at position 0.4 with {\arrow{latex}}, mark=at position 0.7 with {\arrow{latex}},mark=at position 0.90 with {\arrow{latex}}},postaction={decorate}]
    (x1) to[out=180,in=-90] (Rinner) to[out=90,in=180] (x2) to[out=0,in=90] (Router) to[out=-90,in=0] (x1);
        \draw[photon,thick] (x2) -- (r2);
        \draw[photon,thick] (r2) -- (y);
        \draw[photon,gray] (yi) -- (r2);
        \node[right] at (1,1.25) {$\Delta^{(0)}_{\mu\nu}(x_2;\widetilde{y})$};
        \draw[-,dotted,opacity=0.7] (x) -- (xi);
        \draw[-,dotted,opacity=0.7] (y) -- (yi);
    \end{tikzpicture}
    \\ 
    
    \begin{tikzpicture}[scale=0.8]
            \message{Attempt}
        \coordinate (plate_base) at (0,-3);
        \coordinate (plate_top) at (0,3);
        \coordinate (x) at (1,-2);
        \coordinate (y) at (1,2);
        \coordinate (x1) at (1,-0.5);
        \coordinate (x2) at (1,0.5);
        \coordinate (Rinner) at (0.5,0);
        \coordinate (Router) at (1.5,0);
        \coordinate (xi) at (-1,-2);
        \coordinate (yi) at (-1,2);
        \coordinate (r1) at (0,-1.25);
        \coordinate (r2) at (0,1.25);
        \node[above] at (plate_top) {$P$};
        \node[below] at (x) {$x$};
        \node[above] at (y) {$y$};
        \node[below] at (xi) {$\widetilde{x}$};
        \node[above] at (yi) {$\widetilde{y}$};
        \draw[-,gray,thick] (plate_base) -- (plate_top);
        \draw[photon,thick] (x) -- (r1);
        \draw[photon,thick] (r1) -- (x1);
        \node[right] at (1,-1.25) {$\Delta^{(0)}_{\mu\nu}( \widetilde{x};x_1)$};
        \draw[-,black,thick,decoration={markings,mark=at position 0.2 with {\arrow{latex}}, mark=at position 0.4 with {\arrow{latex}}, mark=at position 0.7 with {\arrow{latex}},mark=at position 0.90 with {\arrow{latex}}},postaction={decorate}]
    (x1) to[out=180,in=-90] (Rinner) to[out=90,in=180] (x2) to[out=0,in=90] (Router) to[out=-90,in=0] (x1);
        \draw[photon,thick] (x2) -- (y) node[midway,right] {$\Delta^{(0)}_{\mu\nu}(x_2;y)$};
        \draw[photon,gray] (xi) -- (r1);
        \draw[-,dotted,opacity=0.7] (x) -- (xi);
        \draw[-,dotted,opacity=0.7] (y) -- (yi);
    \end{tikzpicture}
    \qquad 
     \begin{tikzpicture}[scale=0.8]
            \message{Attempt}
        \coordinate (plate_base) at (0,-3);
        \coordinate (plate_top) at (0,3);
        \coordinate (x) at (1,-2);
        \coordinate (y) at (1,2);
        \coordinate (x1) at (1,-0.5);
        \coordinate (x2) at (1,0.5);
        \coordinate (Rinner) at (0.5,0);
        \coordinate (Router) at (1.5,0);
        \coordinate (xi) at (-1,-2);
        \coordinate (yi) at (-1,2);
        \coordinate (r1) at (0,-1.25);
        \coordinate (r2) at (0,1.25);
        \node[above] at (plate_top) {$P$};
        \node[below] at (x) {$x$};
        \node[above] at (y) {$y$};
        \node[below] at (xi) {$\widetilde{x}$};
        \node[above] at (yi) {$\widetilde{y}$};
        \draw[-,gray,thick] (plate_base) -- (plate_top);
        \draw[photon,thick] (x) -- (r1);
        \draw[photon,thick] (r1) -- (x1);
        \node[right] at (1,-1.25) {$\Delta^{(0)}_{\mu\nu}( \widetilde{x};x_1)$};
        \draw[-,thick,black,decoration={markings,mark=at position 0.2 with {\arrow{latex}}, mark=at position 0.4 with {\arrow{latex}}, mark=at position 0.7 with {\arrow{latex}},mark=at position 0.90 with {\arrow{latex}}},postaction={decorate}]
    (x1) to[out=180,in=-90] (Rinner) to[out=90,in=180] (x2) to[out=0,in=90] (Router) to[out=-90,in=0] (x1);
        \draw[photon,thick] (x2) -- (r2);
        \draw[photon,thick] (r2) -- (y);
        \node[right] at (1,1.25) {$\Delta^{(0)}_{\mu\nu}(x_2;\widetilde{y})$};
        \draw[photon,gray] (yi) -- (r2);
        \draw[photon,gray] (xi) -- (r1);
        \draw[-,dotted,opacity=0.7] (x) -- (xi);
        \draw[-,dotted,opacity=0.7] (y) -- (yi);
    \end{tikzpicture}
    \caption{Diagrammatic representation of the $\delta\overline{\Delta}^{(1)}_{\mu\nu}$ correction corresponding to the prescription used in the calculations. Here $P$ is the plate and the gray photon lines represent fictitious propagation from the image coordinates $\widetilde{x},\widetilde{y}$.}
    \label{diagram_prescription}
\end{figure}

Explicitly, we have
\begin{eqnarray}
    \delta\overline{\Delta}^{(1)}_{\mu\nu}(x;y) &=&  \int_\mathcal{D}\dd[4]{x_1}\int_\mathcal{D}\dd[4]{x_2} \Delta^{(0)}_{\mu\alpha}(x-x_1) i\Pi^{\alpha\beta}(x_1 - x_2) \Delta^{(0)}_{\beta\nu}(x_2-y)\nonumber\\
    &&-\int_\mathcal{D}\dd[4]{x_1}\int_\mathcal{D}\dd[4]{x_2} \Delta^{(0)}_{\mu\alpha}(x-x_1) i\Pi^{\alpha\beta}(x_1 - x_2)\Delta^{(0)}_{\beta\nu}(x_2-\widetilde{y})\nonumber\\
    &&-\int_\mathcal{D}\dd[4]{x_1}\int_\mathcal{D}\dd[4]{x_2} \Delta^{(0)}_{\mu\alpha}(\widetilde{x}-x_1) i\Pi^{\alpha\beta}(x_1 - x_2) \Delta^{(0)}_{\beta\nu}(x_2-y)\nonumber\\
    &&+\int_\mathcal{D}\dd[4]{x_1}\int_\mathcal{D}\dd[4]{x_2} \Delta^{(0)}_{\mu\alpha}(\widetilde x-x_1) i\Pi^{\alpha\beta}(x_1 - x_2) \Delta^{(0)}_{\beta\nu}(x_2-\widetilde y) \label{boundarydiagrams:modified1}\,.
\end{eqnarray}
Therefore, one can picture the one loop correction as coming from the four possibilities for the propagation of the photon from $x$ to $y$, which are given by each of the terms in Eq.~(\ref{boundarydiagrams:modified1}). The diagrams with the negative sign correspond to a single reflection, while the ones with positive signs correspond to an  even number of reflections. These terms are depicted in the (position space) diagrams of Fig.~\ref{diagram_prescription}. 

\subsection{Resulting potential}

Eq.~(\ref{1plate:1loop_modifiedpropagator}) can be used to extract the modified 1 loop corrected potential by, once again, invoking Maxwell's equations. Namely, we take
\begin{eqnarray}
    i\overline{A}^{(1)}_\mu(x)  =\int_{\mathbb{R}^{2,1}\times\mathbb{R}_+}\dd[4]{y} \overline{\Delta}^{(1)}_{\mu\nu}(x , y) j^\nu(y)\,,\label{1plate:modified_potential}
\end{eqnarray}
with $j^\nu$ being the source current density. For the Uehling potential, one considers the static charge distribution $j^\nu(y) = q \delta^{\nu}_{\ 0} \delta^3(\vb{y} - \vb{x}_s)$, with $\vb{x}_s = (0,0,L)$ and again only the 0 component, denoted by $\overline{\phi}^{(1)} (x)$ survives.  Using Eq.~(\ref{1plate:1loop_modifiedpropagator}) with expression (\ref{1plate:modified_potential}) and the static charge distribution, the potential can be calculated. 

Before proceeding with the results, it is noteworthy that the static point charge current density allows for integration of $y^0$ which leads to a delta function $\delta(p^0)$. On the support of this delta function,  $\omega_\parallel$ becomes a purely imaginary quantity, namely $\omega_\parallel = \pm i|\vb{p}_\parallel|$.  The positive sign must then be chosen in order to satisfy the boundary condition for $x^\perp_s \to +\infty$. 

The modified corrected potential can be written as
\begin{eqnarray}
    \overline{\phi}^{(1)}(\vb{x}) = \overline{\phi}^{(0)} (\vb{x}) + \delta\overline{\phi}^{(1)} (\vb{x})\,,
\end{eqnarray}
with the loop correction given by
\begin{eqnarray}
    \delta\overline{\phi}^{(1)}(\vb{x})&=&  q\int\frac{\dd[3]{\vb{p}}}{(2\pi)^3}\frac{\Pi(-\vb{p}^2)}{\vb{p}^2}\bigg\{\frac{1}{4}\left[e^{i\vb{p}\cdot(\vb{x} - \vb{x}_s)} - e^{i\vb{p}\cdot(\widetilde{\vb{x}} - \vb{x}_s)} - e^{i\vb{p}\cdot(\vb{x} - \widetilde{\vb{x}}_s)} + e^{i\vb{p}\cdot(\widetilde{\vb{x}} - \widetilde{\vb{x}}_s)}\right] \nonumber\\
    && + \left(e^{-|\vb{p}_{\parallel}|x_s^\perp} - \cos(\vb{p}_\perp\cdot\vb{x}_s) \right)\sin(\vb{p}_\perp\cdot \vb{x}) e^{i\vb{p}_\parallel\cdot(\vb{x} - \vb{x}_s)}\nonumber\\
    && -\left(-e^{-|\vb{p}_\parallel|x^\perp} + \cos(\vb{p}_\perp\cdot\vb{x})\right)\sin(\vb{p}_\perp\cdot\vb{x}_s)e^{i\vb{p}_\parallel\cdot(\vb{x} - \vb{x}_s)} \nonumber\\
    &&\left. + e^{-|\vb{p}_\parallel|(x^\perp + x_s^\perp)}e^{i\vb{p}_\parallel\cdot(\vb{x} - \vb{x}_s)}  + \cos(p^\perp x^\perp)\cos(p^\perp x_s^\perp) e^{i\vb{p}_\parallel\cdot(\vb{x} - \vb{x}_s)}\right.\nonumber\\
    && - \cos(p^\perp x^\perp) e^{-|\vb{p}_\parallel|x_s^\perp} e^{i\vb{p}_\parallel\cdot(\vb{x} - \vb{x}_s)} - e^{-|\vb{p}_\parallel|x^\perp}\cos(p^\perp x_s^\perp) e^{i\vb{p}_\parallel\cdot(\vb{x} - \vb{x}_s)}\bigg\}\,,\label{1plate:potential_correction}
\end{eqnarray}
where $\widetilde{\vb{x}} := (x^1,x^2, -x^3)$. Note that, as a consistency check, $\delta\overline{\phi}^{(1)}$ vanishes for $x^3 = 0$.

Most of the terms in expression (\ref{1plate:potential_correction}) can be integrated. To do so, it is convenient to write $\Pi_R(-\vb{p}^2)$ as in Eq.~(\ref{param_Pi2}), that is
\begin{eqnarray}
    \Pi_R(-\vb{p}^2) = \frac{\alpha}{\pi} \frac{\vb{p}^2}{4m^2}\int_{0}^1 \dd{v} \frac{v^2\left(1 - \frac{1}{3}v^2\right)}{1 + (\vb{p}/2m)^2(1-v^2)}\,.\nonumber
\end{eqnarray}
Since $\Pi_R(-\vb{p}^2)$ is even in the components of the three momentum, the terms proportional to $\sin(p^\perp x^\perp)$ in Eq.~(\ref{1plate:potential_correction}) vanish automatically. The term with the cosine product $\cos(p^\perp x^\perp)\cos(p^\perp x_s^\perp)$ can be expanded into exponentials and combined with the terms in the first line of Eq.~(\ref{1plate:potential_correction}). Thus, the correction is simplified considerably to
\begin{eqnarray}
    \delta\overline{\phi}^{(1)}(\vb x)&=&  q\int\frac{\dd[3]{\vb{p}}}{(2\pi)^3}\frac{\Pi(-\vb{p}^2)}{\vb{p}^2}\left\{e^{i\vb{p}\cdot(\vb{x} - \vb{x}_s)}  +  e^{-|\vb{p}_\parallel|(x^\perp + x_s^\perp)}e^{i\vb{p}_\parallel\cdot(\vb{x} - \vb{x}_s)}\right.\nonumber\\
    &&\left. - \cos(p^\perp x^\perp) e^{-|\vb{p}_\parallel|x_s^\perp} e^{i\vb{p}_\parallel\cdot(\vb{x} - \vb{x}_s)} - e^{-|\vb{p}_\parallel|x^\perp}\cos(p^\perp x_s^\perp) e^{i\vb{p}_\parallel\cdot(\vb{x} - \vb{x}_s)}\right\}.\label{1plate_deltaA2}
\end{eqnarray}


As expected, in the limit $x^\perp_s \to \infty$ with  $\mathbf{x}-\mathbf{x}_s$ fixed, we re-obtain the result in the absence of the plate given by Eq.~(\ref{Eq10}).    Moreover, when $x^\perp_s\to0$ we obtain $\delta\overline{\phi}^{(1)}(\vb x)\to 0$, as it should. In this limit the grounded plate shields away completely any field produced by the point charge.

Note that the influence of the plate is given by
the last three terms in Eq.~(\ref{1plate_deltaA2})---precisely the terms which vanish in the limit $x_s^\perp \to \infty$. These terms cannot easily be  analytically integrated. The term containing $\exp(-|\vb{p}_\parallel|(x^\perp + x_s^\perp))$, for example, can be written as
\begin{eqnarray}
      \frac{\alpha m}{2\pi^2} \int_0^1 \frac{v^2(1-v^2/3)}{(1 -v^2)^{3/2}} dv\int_{0}^{\infty} \frac{\rho d\rho}{\sqrt{1 + \rho^2}}  J_0(a(v)\rho ) e^{-b(v) \rho }\,,
      \label{1plate:sample_potentialform}
\end{eqnarray}
where $J_0(x)$ is the zeroth order Bessel function of the first kind and the other functions above are defined as
\begin{eqnarray}
    a(v) :=  2m\frac{|\vb{x}^\parallel - \vb{x}_{s}^{\parallel}|}{\sqrt{1-v^2}}\,, \qquad\qquad b(v):= 2m\frac{x^\perp + x_{s}^\perp}{\sqrt{1-v^2}}\,.\nonumber
\end{eqnarray}

In Appendix \ref{App_B}, we show a possible expression for (\ref{1plate:sample_potentialform}) in terms of special functions. Here, instead, we will express all three contributions in a form closely related  to the standard Uehling potential by changing variables to $\xi = (1 - v^2)^{-1/2}$. Thus, the correction $\delta\overline{\phi}^{(1)}(\vb{x})$ becomes
\begin{eqnarray}
    \delta\overline{\phi}^{(1)}(\vb{x}) &=& \frac{q\alpha}{6\pi^2}\int_1^\infty \dd{\xi}\left(1 + \frac{1}{2\xi^2}\right)\frac{\sqrt{\xi^2 - 1}}{\xi^2}\bigg\{ \frac{e^{-2m\xi|\vb{x}-\vb{x}_s|}}{|\vb{x} - \vb{x}_s|} \nonumber\\
    &&+ \int_0^\infty \frac{\rho J_0(\rho|\vb{x}^\parallel-\vb{x}_{s}^{\parallel}|)\dd{\rho}}{\sqrt{\rho^2 + 4m^2 \xi^2}} \nonumber\\
    &&\times\left[ e^{-\rho(x^\perp + x_{s}^{\perp})} - e^{-(\rho x^\perp + x_{s}^{\perp}\sqrt{\rho^2 + 4m^2\xi^2})}- e^{-(\rho x_{s}^{\perp} + x^{\perp}\sqrt{\rho^2 + 4m^2\xi^2})}\right]\bigg\}\,,
    \label{1plate:loop_correction_expanded}
\end{eqnarray}
and we explicitly see that the first term above is exactly the term appearing in the standard form of the Uehling correction---see Eq.~(\ref{UehlingPotential1}). 

In order to further simplify the notation, we will label the integral appearing in (\ref{UehlingPotential1}) as $\mathcal{I}(r)$ ($r:=|\vb{x}-\vb{x}_{s}|$) and the double integral appearing in Eq.~(\ref{1plate:loop_correction_expanded}) as $\mathcal{F}(\vb{x},\vb{x}_s)$. Specifically,
\begin{eqnarray}
    \mathcal{I}(r) &:=& \int_1^\infty\dd{\xi}\left(1 + \frac{1}{2\xi^2}\right) \frac{\sqrt{\xi^2 - 1}}{\xi^2} e^{-2 m \xi r}\,,\nonumber\\
    \mathcal{F}(\vb{x},\vb{x}_s) &:=& \int_1^\infty \dd{\xi}\left(1 + \frac{1}{2\xi^2}\right)\frac{\sqrt{\xi^2 - 1}}{\xi^2} \int_0^\infty \frac{\rho J_0(\rho |\vb{x}^\parallel-\vb{x}^\parallel_{s}|)\dd{\rho}}{\sqrt{\rho^2 + 4m^2 \xi^2}}\nonumber\\
    &&\times\left[ e^{-\rho(x^\perp + x_{s}^{\perp})} - e^{-(\rho x^\perp + x_{s}^{\perp}\sqrt{\rho^2 + 4m^2\xi^2})}  - e^{-(\rho x_{s}^{\perp} + x^{\perp}\sqrt{\rho^2 + 4m^2\xi^2})}\right].\nonumber
\end{eqnarray}
In terms of these two functions, the ${\cal{O}}(\alpha)$ correction in the presence of the plate is expressed as
\begin{eqnarray}
    \delta\overline{\phi}^{(1)}(\vb{x}) = \frac{q \alpha}{6\pi^2 r}\left[\mathcal{I}(r) + r\mathcal{F}(\vb{x},\vb    x_s)\right].\label{1plate_deltaA_final}
\end{eqnarray}
The full potential is then expressed as the sum of the tree-level contribution and the 1 loop modification, i.e.
\begin{eqnarray}
    \overline{\phi}^{(1)}(\vb{x}) &=& \frac{q}{4\pi}\bigg\{\frac{1}{r} - \frac{1}{\widetilde{r}}+ \frac{2\alpha}{3\pi}\bigg[\frac{\mathcal{I}(r)}{r} + \mathcal{F}(\vb{x},\vb{x}_s)\bigg]\bigg\},\,
    \label{1plate_final_result}
\end{eqnarray}
with  $\widetilde{r} :=|\vb{x} - \widetilde{\vb{x}}_{s}|$, and we recall that $\widetilde{\vb{x}}_{s} = \vb{x}^\parallel_s - \vb
x^\perp_s$ (which corresponds to the image charge position). 

\section{Numerical results}\label{section:results}

\begin{figure}[!h]
    \centering
    \includegraphics[width=0.6\linewidth]{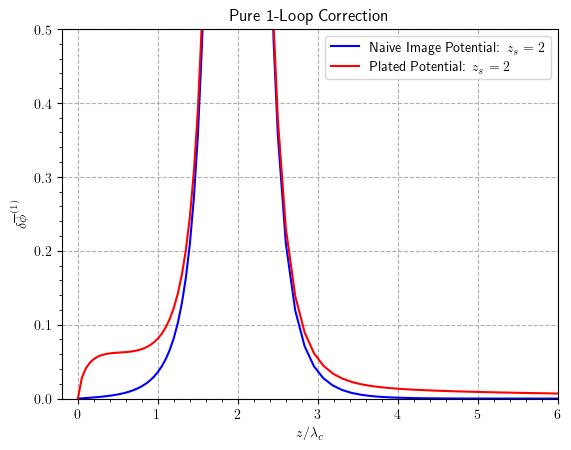}
    \caption{Plot of the linear-in-$\alpha$ correction to Coulomb's law for a charge located at $\vb{x}_s = (0,0,2\lambda_c)$, considering the potential found in expression (\ref{1plate_deltaA_final}) (red) and the naive prescription of the method of images (blue). Here we set $m=1$ for the fermion mass in the loop.}
    \label{fig1:1plate_correction}
\end{figure}

From Eq.~(\ref{1plate_final_result}), we see that the classical interpretation of the potential from the method of images as a superposition of contributions from the charge and its image is lost at the quantum level (already at the first loop order). This distinction can be explicitly seen in Fig.~\ref{fig1:1plate_correction}, where the linear-in-$\alpha$ correction obtained in Eq.~(\ref{1plate_deltaA_final}) is plotted alongside the  potential which would be obtained from  a naive application of the method of images, given by $  \delta\overline{\phi}^{(1)\rm naive}(\vb{x}) = \frac{q \alpha}{6\pi^2 }(\mathcal{I}(r)/r -\mathcal{I}(\widetilde{r})/\widetilde{r})$. We see  that the nonlinearity enhances the Uehling correction for all distances. In particular, the Uehling correction rises much more steeply near the plate than what would follow from a direct image superposition.



\begin{figure}[!h]
    \centering
    \includegraphics[width=0.6\linewidth]{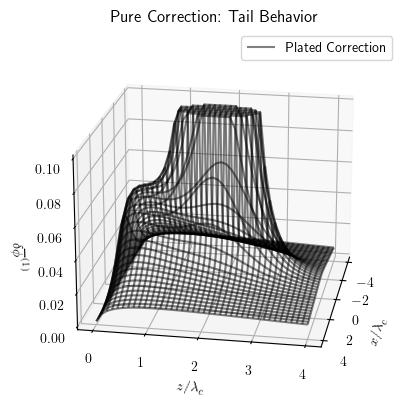}
    \caption{Three-dimensional plot of numerical samples obtained for the pure loop correction term, $\delta\overline{\phi}^{(1)}$, with a charge placed at $\vb{x}_s = (0,0,2\lambda_c)$. Once again, we set $m=1$.}
    \label{fig2:1plate_3d1}
\end{figure}

\begin{figure}[!h]
    \centering
    \includegraphics[width=0.6\linewidth]{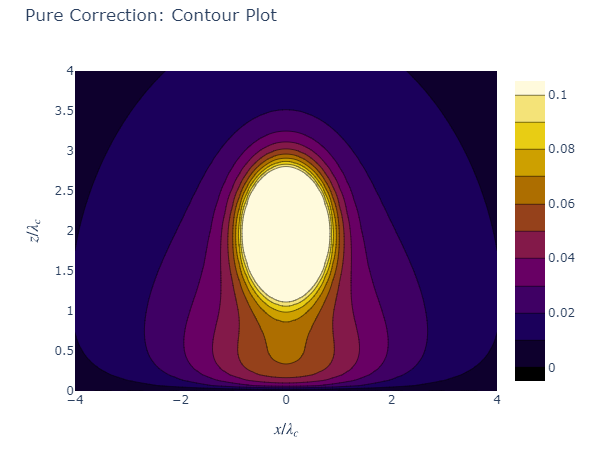}
    \caption{Contour plot of the pure loop correction term, $\delta\overline{\phi}^{(1)}$, with a charge placed at $\vb{x}_s = (0,0,2\lambda_c)$. Once again, we set $m=1$.}
    \label{fig4:1plate_density}
\end{figure}

\begin{figure}[h]
    \centering
    \includegraphics[width=0.6\linewidth]{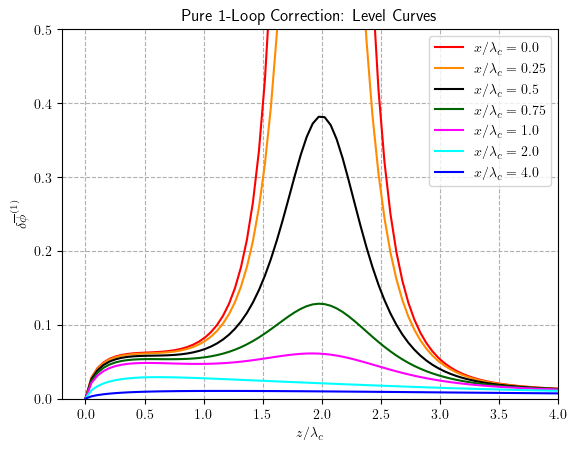}
    \caption{Transversal profile of the pure loop correction term, $\delta\overline{\phi}^{(1)}$, with a charge placed at $\vb{x}_s = (0,0,2\lambda_c)$. Once again, we set $m=1$.}
    \label{fig3:1plate_crosssectional}
\end{figure}

In Fig.~\ref{fig2:1plate_3d1} we present a 3-dimensional plot of the Uehling potential at points $(x,0,z)$,  for the case of a point charge located at $(0,0,2\lambda_c)$, where $\lambda_c$ denotes the Compton wavelength of the fermion. Note that, given the azimuthal  symmetry of the configuration, the dependence on $y$ must be the same as that on $x$. Equivalently, Fig.~\ref{fig4:1plate_density} represents the same situation in a contour plot, while Fig.~\ref{fig3:1plate_crosssectional} presents the Uehling potential as a function of the distance of the plate as a function of $x$ (corresponding to cross sections of the plots given in Fig.~\ref{fig2:1plate_3d1}. 
Again we notice a great enhancement of the Uehling potential near the plate. 

\begin{figure}[!h]
    \centering
    \includegraphics[width=0.6\linewidth]{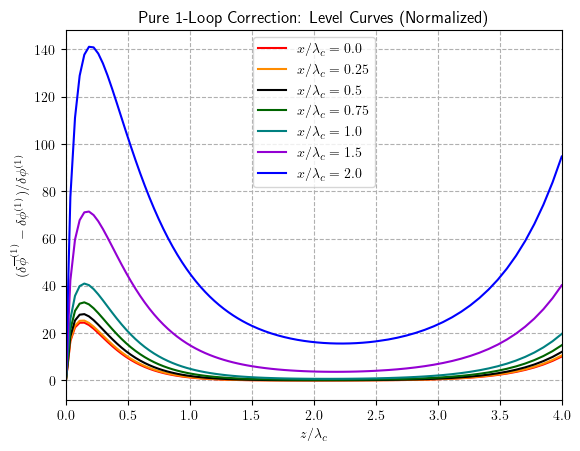}
    \caption{Plot of the normalized linear-in-$\alpha$ contribution considering the presence of the plate ($\delta\overline{\phi}^{(1)}$), normalized by the standard potential without the plate ($\delta\phi^{(1)}$), with $\vb{x}_s = (0,0,2\lambda_c)$. Also, we set $\alpha = 1/137$, $q = 1$, $m=1$, as before.}
    \label{fig5:1plate_normalized_potential_uehling}
\end{figure}

\begin{figure}[!h]
    \centering
    \includegraphics[width=0.6\linewidth]{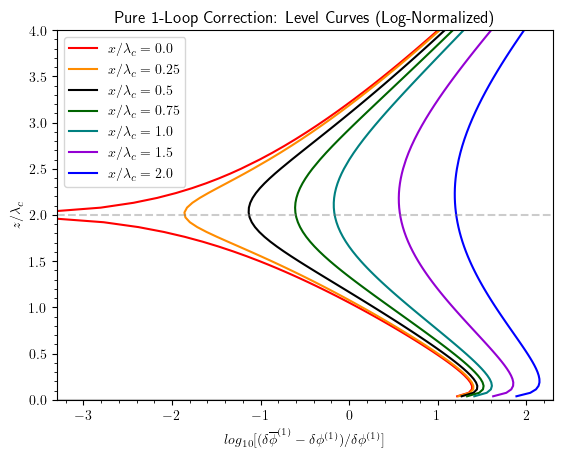}
    \caption{Log scale plot of the normalized linear-in-$\alpha$ contribution considering the presence of the plate ($\delta\overline{\phi}^{(1)}$), normalized by the standard potential without the plate ($\delta \phi^{(1)}$), with $\vb{x}_s = (0,0,2\lambda_c)$. Also, we set $\alpha = 1/137$, $q = 1$, $m=1$, as before.}
    \label{fig7:1plate_log_normalized_potential_uehling}
\end{figure}

Consider now varying $z$ for a fixed value of $x$, starting from $z=0$ (i.e. the plate) and increasing $z$.  At $z=0$ the Uehling correction vanishes for all values of $x$, as   anticipated. As we increase $z$ the Uehling correction rises to a maximum value at a position $z_m(x)$ and then decreases monotonically.  The point where the distance to the charge position is the smallest is at $z=2\lambda_c$, but note that except for $x=0$ (where the correction diverges at $z=2\lambda_c$) we have $z_m(x)<2\lambda_c$, showing that the maximum Uehling correction for a given $x$ is shifted towards the plate. 
In order to isolate the effect of the plate in Fig.~\ref{fig5:1plate_normalized_potential_uehling} we plot the difference between the Uehling correction with and without the plate, normalized by the latter. We see that the plate can increase the effect by several orders of magnitude. This can also be observed in the log-scale plot of Fig.~\ref{fig7:1plate_log_normalized_potential_uehling}, which displays clearly the shift of the maximum towards the plate as we vary $x$. 

\section{Conclusions}\label{secion:conclusion}


\

In this work, we have explicitly calculated the influence of a perfectly conducting plate on the first loop correction of the Coulomb potential (the Uehling correction). For this, we have obtained the one loop correction corresponding to vacuum polarization for the photon propagator subject to the Dirichlet boundary condition imposed by the plate.
We have shown that the Uehling correction is not given by the naive image superposition due to non-linearity effects which strongly enhance the Uehling correction near the plate. In addition, the Uehling correction can be augmented by orders of magnitude at some positions in comparison with the case without the $z=0$ boundary.       

The study of boundary effects on the Uehling potential opens new avenues of work. A first perspective is to analyze the boundary induced corrections when the point charge is between a pair of parallel plates. Here the infinite images and the non-additivity correction could yield an even stronger modification to the electrostatic potential, leveraging the possibility of experimental verification at distances from the charge where usually the correction is undetectable.
Also interesting, albeit less straightforward, are the higher order corrections, which could benefit from the recent developments in loop calculations based on on-shell methods \cite{Bern_2011,Ellis_2012,Bern_1994_oneloop_npoint}. 
Another interesting development is performing the analogous calculation in other theories, such as scalar QED. One reason these other cases could be worth investigating stems from the general current interest in exploring interconnections between different theories, which often allows one to obtain  results for one theory by doing calculations in another (simpler) theory, for example through the double-copy paradigm \cite{bern2008new,bern2010perturbative,bern2024sagexreviewscatteringamplitudes,bern2022scalar}.  


\begin{acknowledgments}

The work of TA was partially supported by the European Structural and Investment Funds and the Czech Ministry of Education, Youth and Sports (project
{\tt FORTE CZ.02.01.01/00/\\ 22\_008/0004632}), through its research mobility program. The work of G.Z. was supported by the Coordenação de Aperfeiçoamento de Pessoal de Nível Superior (CAPES). C.F. acknowledges funding from CNPq (Grants No. 308641/2022-1 and 408735/2023-6) and FAPERJ (Grant No. 204.376/2024). F.A.B thanks CNPq under the grant 313426/2021-0.

\end{acknowledgments}

\appendix

\section{Restricted Domain Integral}\label{App_A}

As discussed in the text, evaluation of the half-plane restricted integrals appearing in Eq.~(\ref{1plate:modified_loop_contribution1}) naturally requires the use of the Sokhotski-Plemelj theorem. This is because the oscillating integrals must be regularized. Explicitly, with $\mathcal{D}:=\mathbb{R}^{2,1} \times \mathbb{R}^+$,

\begin{eqnarray}
    \int_{\mathcal{D}}\dd[4]{x_2} e^{ix_2\cdot(p_2 - p_3 - p_4)} &=& \int_\mathbb{R}\dd{x_2^0} e^{ix_2^0(p_2^0 - p_3^0 - p_4^0)}\int_{\mathbb{R}^2} \dd{\vb{x}_{2\parallel}} e^{i\vb{x}_{2\parallel}\cdot(\vb{p}_{4\parallel} + \vb{p}_{3\parallel} - \vb{p}_{2\parallel})} \nonumber\\
    &&\times\int_0^{\infty} \dd{x_2^3} e^{ix^3_2 (p^3_4 + p^3_3 - p_2^3)} \nonumber\\
    &=& (2\pi)^4\delta(p_2^0 - p_3^0 - p_4^0)\delta^2(\vb{p}_{4\parallel} + \vb{p}_{3\parallel} - \vb{p}_{2\parallel}) \nonumber\\
    &&\times \left[ \frac{1}{2}\delta(p^3_4 + p_3^3 - p_2^3)+ \frac{i}{2\pi}\mathcal{P}\frac{1}{p^3_4 + p_3^3 - p_2^3}\right]\,,
\end{eqnarray}
where $\mathcal{P}$ refers to the Cauchy principal value. Thus, the two vertex integrals in (\ref{1plate:modified_loop_contribution1}) lead to four terms which  are given by Eqs. (\ref{1plate:Omega1}), (\ref{1plate:Omega2.1}), (\ref{1plate:Omega3.1}) and (\ref{1plate:Omega_4.1}).

\section{Closed form of Eq.~(\ref{1plate:sample_potentialform})}\label{App_B}

As advertised in the discussion of the modified 1 loop corrected potential term $\delta\overline{A}_\mu$, the integral of the Bessel function appearing in Eq.~(\ref{1plate:sample_potentialform}) can be expressed analytically in terms of other special functions. By labeling $A:=2m\xi|\vb{x}_\parallel - \vb{x}_{e\parallel}|, B:= 2m\xi(x_\perp + x_{e\perp})$, the integral reads
\begin{eqnarray}
    I:=\int_0^\infty \frac{\rho}{\sqrt{1+\rho^2}} J_0(A\rho) e^{-B\rho} \dd{\rho}\,.
\end{eqnarray}
Identifying the inverse square root above as the Laplace transform of $J_0(\rho)$, the integral can be written as
\begin{align}
     I=\int_0^\infty\dd{\tau} J_0(\tau) \int_0^\infty\dd{\rho} \rho J_0(\rho A) e^{-\rho(B + \tau)}\,.\nonumber
\end{align}
Using Eq.~(WA422(6)) from reference \cite{gradshteyn}, the above $\dd\rho$ integral can be performed, yielding
\begin{align}
     I=\int_0^\infty \frac{J_0(\tau) (\tau + B)}{(A^2 + (\tau + B)^2)^{3/2}}\dd{\tau}\,.
\end{align}
At last, with the aid of the \textsc{Mathematica} software, the above can be explicitly written in terms of multiple special functions as below, 
\begin{align}
    I =&\, -\pdv{B}\int_0^\infty \frac{J_0(\tau)}{\sqrt{A^2 + (B + \tau)^2}}\dd{\tau}\nonumber\\
    =&\, \frac{1}{2iA} \left(i (A+i B) {}_0F_1^{(1,1)}\left(1,\frac{1}{4} (A+i B)^2\right)+(B+i A) {}_0F_1^{(1,1)}\left(1,\frac{1}{4} (A-i B)^2\right)\right.\nonumber\\
    &\,-\pi  \pmb{H}_{-1}(i A-B)+\pi  \pmb{H}_{-1}(B+A i) +\frac{2 I_0(A+i B )}{B-i A}-\frac{2 I_0(A-i B)}{B+i A}\nonumber\\
    &\,-\ln2 \left(J_1(i A-B)+J_1(B+ iA)\right) \nonumber\\
    &\left.+2 i \left(\ln\left(B-i A\right) I_1(A+iB ) + \ln \left(B+ i A\right) I_1(A-i B)\right)\right)\,.
    \label{appendix:ugly1}
\end{align}
Here $_0F_1$ is the confluent hypergeometric function, $\pmb{H}_0$ is the Struve function of first kind, $I_\nu$ is the modified Bessel function of the second kind and $J_\nu$ is the Bessel function of the first kind. As we noted in the text, the above is not very enlightening, justifying our choice to resort to the numerical analysis.

\bibliographystyle{JHEP}
\bibliography{cites}

\end{document}